# *The structure of technological learning: insights from water electrolysis for cost forecasting, policy, and strategy*

Mohamed Atouife[1,*] and Jesse Jenkins[1,*]

[1] Mechanical and Aerospace Engineering Department, Princeton University, Princeton, New Jersey, 08540, USA.

[*] Correspondence: m.atouife@princeton.edu, jdj2@princeton.edu

## Summary

Forecasting the cost evolution of emerging clean technologies is crucial for informed policy, investment, and decarbonization decisions, yet it remains deeply uncertain. Learning curves, which link cost declines to cumulative deployment, are widely used for technological cost forecasting. However, applying them to emerging technologies is challenging due to parametric uncertainty in learning rates, which are scarce and highly uncertain, and structural uncertainty stemming from multiple plausible learning frameworks. Using water electrolysis as a case study, we evaluate how different learning structures, from shared to fragmented learning across technology variants and regions, alter expected cost paths. We interrogate model assumptions that represent contrasting industrial realities, including competition among electrolyzer variants and supply chain fragmentation associated with protectionism and industrial policy. We find that plausible modeling choices generate widely different trajectories, with materially different implications for policy design and technology strategy. We argue for routinely applying multiple learning frameworks to explore decision spaces and stress-test conclusions for scale-up planning, national industrial strategy, and energy-systems modeling.

## Introduction

Experience curves, often referred to as Wright's law[1] or learning curves, are widely used to analyze the cost evolution of energy technologies[2–9]. In their simplest form, they relate declines in unit costs to cumulative deployment, based on the premise that accumulated production and deployment experience lowers costs over time[10].

Because experience curves are often used to project the future costs of emerging clean technologies[3,11], they play an important role in shaping policy and investment expectations[12]. These projections inform judgments about when a technology may become competitive, how much early support may be required, and to what extent near-term deployment can drive future cost reductions. The usefulness of such projections therefore depends on whether their assumptions adequately reflect how costs are likely to evolve.

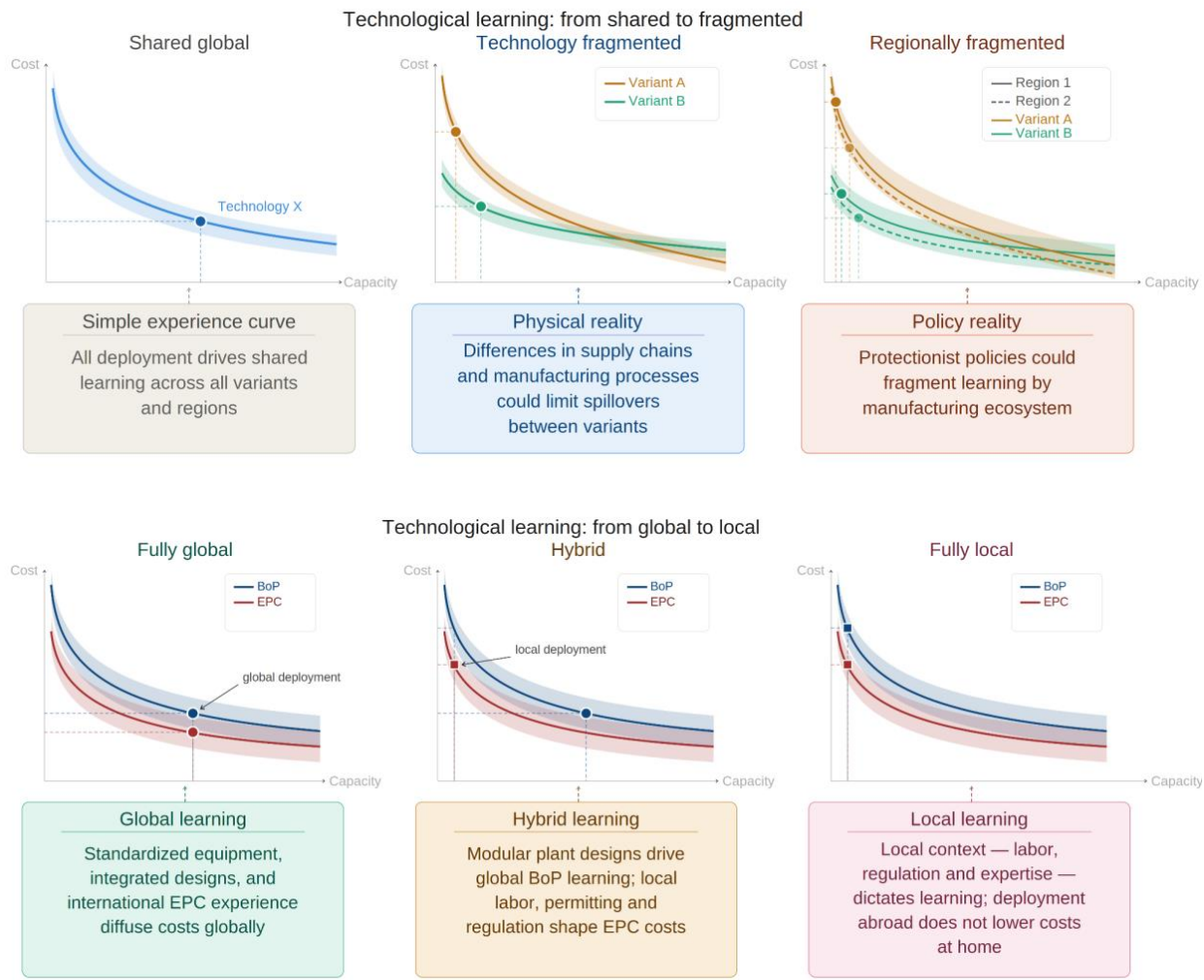

**Graphical Abstract: Stylized learning trajectories under alternative fragmentation structures.** *Top row:* Variant A and B represent two technological variants of Technology X (e.g. alkaline and PEM electrolyzers). Region 1 and 2 represent distinct manufacturing ecosystems (e.g. Western and Chinese). *Bottom row:* BoP (Balance of Plant) and EPC (Engineering, Procurement and Construction) are the two main non-stack cost components of a Technology X project. Learning rates and deployment levels are illustrative.

While the simplicity of the experience curve framework contributes to its appeal, its application to emerging technologies necessitates careful consideration. Although empirical evidence strongly supports learning effects in specific technologies, such as solar PV modules[13], these effects are not universally observed. Nuclear reactors, for example, have not experienced similar cost declines and have instead exhibited instances of negative learning[14], with costs increasing over time globally. However, evidence from South Korea suggests that successful localized learning is possible[15]. Furthermore, the standard experience curve model typically attributes cost reductions solely to cumulative deployment, thereby treating deployment as an all-encompassing proxy and overlooking the multiple mechanisms that can drive cost declines.

In practice, cost reductions arise through several mechanisms, including economies of scale in manufacturing, reduced capital costs as technological risk declines, research-driven innovation, advances in plant and manufacturing design, and growing labor expertise in installation and operation[16–18]. Different components of clean energy technologies may therefore experience cost reductions through distinct mechanisms. For example, cost declines for solar PV modules are predominantly driven by innovation in design, automation, and economies of scale in manufacturing, whereas cost reductions for balance-of-system components are more strongly influenced by advances in system design and accumulated installation expertise[16,19]. This distinction can affect both the rate and the nature of cost reductions, including whether learning follows a global or local trajectory[14]. Mass-manufactured technologies, such as lithium-ion battery packs, benefit from global-scale learning effects because their cost reductions are largely attributable to innovation and manufacturing efficiencies independent of final deployment location[16]. Conversely, cost reductions associated with procurement, installation, and other site-specific factors depend more strongly on regional conditions, typically leading to local learning effects shaped by workforce expertise, local competition, and supply chain development[20].

These complexities highlight the challenges involved in applying experience curve models to predict the path of cost reductions for emerging technologies. Selecting an appropriate experience model therefore requires understanding the underlying drivers of cost reduction, which vary across technology types. Furthermore, uncertainty arises not only from estimating key parameters, such as the learning rate (i.e., the percentage reduction in cost with each doubling of installed capacity), which either cannot be empirically determined for nascent technologies or their estimation is associated with great uncertainty, but also from the structural assumptions inherent in the model itself[21]. Specific uncertainties include the distribution of learning effects across different technological components or stages of manufacture/installation, the degree to which learning occurs on a local versus global scale, and whether different variants of the same technology are assumed to experience shared learning. These challenges are further compounded by new realities of increasing protectionism and localization of supply chains.

In this study, we examine different learning model structures for emerging technologies, their underlying assumptions, and the extent to which these modeling choices substantively influence the conclusions and policy implications drawn from experience curve analyses. We use water electrolysis as a case study because it is both an important emerging technology for decarbonization and one for which future cost reductions are widely expected to arise through learning[22]. There is now a growing literature that either estimates experience-curve learning rates for electrolysis and/or uses them to project future cost declines. Some studies treat electrolysis as a single umbrella technology[23], while others focus on a single variant or distinguish between alkaline and proton exchange membrane (PEM) systems[24–28]. The aforementioned studies apply learning curves to total capital costs, whereas others separate stack and non-stack components[29]. Across these different approaches, however, learning is generally treated as global in scope.

Even when studies distinguish between ALK and PEM or adopt a component-based framework, they typically do not examine whether non-stack costs may follow more local learning dynamics or whether supply-chain fragmentation and protectionism may generate distinct regional learning trajectories. These are not neutral modeling choices: they embed assumptions about spillovers across technology variants, the geographic scope of learning, and the extent to which different cost components are expected to evolve together. Using electrolysis, we show how alternative learning structures affect the future cost of electrolytic hydrogen and the magnitude of cumulative deployment and investment required to achieve given cost targets. We also discuss the implications of these alternative learning models for technology policy, industrial policy, and climate mitigation efforts. Because the appropriate learning rate and model structure remain highly uncertain, and multiple defensible choices may coexist, we additionally introduce an online dashboard that allows users to explore these assumptions and their implications transparently[30].

# Results

## Alternative learning models

Water electrolysis project capital costs can be divided into three main components. The first is the electrolyzer stack, which is available in two principal commercial technologies: alkaline (ALK) and proton exchange membrane (PEM). The second is the balance of plant (BoP), which includes supporting infrastructure such as water separation units, compressors, alkaline solution tanks, and other auxiliary systems. The third is engineering, procurement, and construction (EPC), which includes plant design, permitting, and on-site construction and installation.

As noted above, different cost components within a given technology may experience cost reductions at different rates and through different underlying mechanisms. Treating electrolysis as a single aggregate capital cost category risks obscuring these differences and imposing a common learning structure on components that may evolve differently. We therefore adopt a component-based experience curve framework to project future cost trajectories for water electrolysis projects[31]. We begin with electrolyzer stack costs.

## Stack costs

### *Technological fragmentation: one technology, different variants*

Water electrolysis is an umbrella term encompassing multiple technological approaches aimed at splitting water into hydrogen and oxygen using electricity. A crucial methodological question in experience curve modeling is whether different electrolyzer stack variants follow a *shared* learning curve—benefiting from cumulative knowledge and production efficiencies that lower costs for both technologies simultaneously—or whether their cost trajectories evolve independently, leading to *fragmented* learning curves. More specifically, it is important to assess whether ALK and PEM electrolyzers compete for learning opportunities, such that the deployment of one variant does not generate spillover benefits for the other.

ALK and PEM electrolyzer stacks are both mass-manufactured, modular technologies, suggesting that cost reductions primarily originate from advances in manufacturing processes and supply chain development[16]. Therefore, there is potential for positive spillover between the two technology variants to the extent that they share a common manufacturing platform and/or supply chains. While ALK and PEM electrolyzers serve the same fundamental function, they exhibit substantial differences in material composition, design, and production requirements[32]. ALK electrolyzers employ a liquid potassium hydroxide electrolyte and are primarily constructed from relatively low-cost materials such as nickel. In contrast, PEM electrolyzers rely on a membrane and require scarce and expensive materials, including platinum and iridium. These material and design distinctions result in divergent manufacturing processes and supply chain structures. For instance, the production of PEM electrolyzers necessitates highly automated and specialized fabrication techniques, such as precision coating of membranes with thin layers of platinum and iridium—processes that do not apply to ALK electrolyzers[33]. Furthermore, due to their dependence on rare earth metals, the supply chains for ALK and PEM electrolyzers function independently, each requiring sufficient market demand to achieve economies of scale and operational efficiency.

These distinctions also matter for innovation. Historical evidence from the solar module industry suggests that firms establish research and development efforts, and finance innovation and cost reduction, through revenues generated from product sales[34]. On that logic, efforts to reduce reliance on rare and costly materials in PEM electrolyzers would be unlikely in the absence of sustained deployment specifically for PEM stacks. In practice, spillovers between ALK and PEM may exist, but they are unlikely to be complete. The extent to which learning should be treated as shared across the two technologies is therefore a substantive structural assumption.

To illustrate the implications of this assumption, consider a hypothetical scenario in which only ALK electrolyzers are deployed, resulting in no demand for PEM electrolyzers. Under a shared learning curve, widespread deployment of ALK could still be assumed to drive cost reductions in PEM. However, that outcome would embed a strong assumption about spillovers, since the absence of PEM demand would preclude the establishment of PEM production facilities and the associated learning effects. This example highlights why the degree of shared learning between technology variants should be treated as an explicit modeling choice when projecting future stack costs.

### *Regional fragmentation: fragmented world, fragmented learning*

The fragmentation of technological learning may extend beyond the divergence of technology variants following separate learning curves. The revival of industrial policy, the increasing prevalence of protectionist policies, and an emphasis on localizing supply chains and expanding domestic manufacturing have the potential to exacerbate this fragmentation. Previous studies examined the potential impact of localized solar supply chains to result in distinct learning trajectories for solar PV modules manufactured in China and those produced elsewhere in the world[35]. We postulate that a similar dynamic may emerge for mass-manufactured emerging technologies such as water electrolysis. For example, current market trends indicate that Chinese-manufactured ALK stacks are significantly less expensive than their Western counterparts, whereas PEM stacks produced in China and the West currently exhibit relatively similar cost structures[36]. This cost disparity could create a reinforcing cycle in which demand gravitates toward the lower-cost Chinese ALK stacks, driving further cost reductions through increased deployment, while Western-manufactured stacks risk stagnation.

Given that cost reductions in mass-manufactured technologies predominantly result from incremental innovations in production processes and product design—such as minimizing reliance on expensive raw materials or improving manufacturing yields—the degree to which knowledge is shared between Western and Chinese manufacturers will be a key determinant of future cost trajectories. Factors influencing this knowledge exchange include the free flow of technical expertise and tacit knowledge, mobility of employees across firms and nations, collaboration through joint ventures and technology licensing agreements, transfer of know-how via manufacturing equipment providers, and cross-border investments in manufacturing infrastructure. Historical precedent suggests that such mechanisms played a crucial role in the cost reductions observed in the solar industry during the early 2000s when knowledge transfer between Western and Chinese firms facilitated rapid innovation before Chinese firms ultimately gained a dominant position in cost and technological leadership[17].

Grouping all Western manufacturers under a single "Western" category is a stylized simplification that presumes minimal barriers to innovation diffusion and knowledge exchange across Western economies. This assumption suggests that firms are part of one knowledge ecosystem and can operate across multiple markets and share technological advancements. To the degree that policies emerge that restrict international collaboration—such as measures discouraging joint ventures, limiting exports of manufacturing equipment, restricting immigration, or preventing foreign direct investment in manufacturing facilities—supply chains may develop in parallel, leading to further fragmentation of learning. Under such conditions, technological improvements achieved in one region would not yield benefits for others, thereby slowing the overall learning process.

This additional layer of fragmentation introduces the possibility that not only distinct technology variants (ALK vs. PEM) but also their regional subvariants (Western vs. Chinese) could follow separate learning curves, resulting in four independent cost trajectories. This hypothesis provides a conceptual framework for assessing the potential cost implications of protectionist policies in contrast to globally integrated supply chains. The standard experience curve model for water electrolysis, which assumes a unified learning trajectory, may, therefore, require revision to account for these geopolitical and economic factors.

## Balance of Plant and Engineering, Procurement, and Construction Costs

*Local versus global learning*

While modeling electrolyzer stack costs accurately is important, the treatment of balance of plant (BoP) and engineering, procurement, and construction (EPC) costs is arguably even more consequential, since these components account for the majority of capital expenditures in water electrolysis projects. Considerable uncertainty remains regarding the learning rates associated with these cost categories. Experience from the solar industry suggests that balance-of-system costs often exhibit slower learning than core equipment such as modules[14]. As a simplifying assumption, we treat learning in BoP and EPC costs as shared across projects employing ALK and PEM stacks, on the premise that some knowledge gained through plant design optimization for one stack type may carry over to the other. However, this assumption may break down if the engineering requirements or system configurations of ALK and PEM diverge sufficiently, in which case learning could also become fragmented across stack technologies.

A more fundamental question is whether learning in BoP and EPC costs is global or local. This distinction has important implications for projected cost trajectories, industrial strategy, and international competitiveness. If learning is global, cost reductions are driven by cumulative worldwide investment, allowing countries or regions to benefit from declining costs even if they do not bear the initial deployment burden themselves. This dynamic resembles the solar PV module industry, where early adopters such as Germany absorbed high initial costs and later adopters benefited from the resulting cost reductions[17]. If learning is local instead, cost declines are tied to deployment experience gained within a given region. In that case, early movers may gain a lasting advantage by unlocking cost reductions before others accumulate comparable experience. This distinction is particularly important for electrolysis because it affects not only project costs and the competitiveness of hydrogen-derived products such as green ammonia and steel, but also broader climate mitigation pathways. If learning is global, early investment in wealthier countries may help drive down costs and enable later adoption in developing regions. If learning is local, however, those cost reductions may not diffuse as broadly, and the process of capability formation, cost reduction, and deployment may need to be repeated across regions through localized investment and experience. The choice between local and global learning therefore shapes not only industrial advantage, but also the extent to which scale-up in one part of the world can support decarbonization elsewhere.

A defensible starting point is that BoP and EPC learning may be predominantly local. These costs are closely tied to project delivery and are shaped by jurisdiction-specific factors such as labor markets, permitting processes, regulatory requirements, and the degree of competition among local firms. Such conditions vary substantially across regions, making it plausible that cost reductions occur primarily through experience gained within individual markets rather than through broad international spillovers. This pattern is consistent with the case of residential rooftop solar, where soft costs in the United States remain significantly higher than in countries such as Germany or Australia[19].

At the same time, a more global view is also defensible, particularly for BoP. BoP cost reductions can arise from economies of scale in larger plants, standardization and improvements in engineering design, system integration, and plant layout, such as multiple stacks sharing common BoP components, as well as from vertical integration of suppliers. Some providers offer modular turnkey systems that integrate stacks and BoP[37], reducing the need for customization regardless of deployment location and making BoP learning more likely to diffuse internationally. Learning may also globalize through tacit knowledge transferred by international developers and engineering firms operating across jurisdictions, or through the export of EPC-related services by experienced firms[38].

Multiple opposing views of BoP and EPC learning are therefore defensible. One emphasizes the local nature of project delivery and the importance of jurisdiction-specific conditions. The other emphasizes the potential for standardization, engineering optimization, and international project-development experience to diffuse learning across markets. The case for broader international spillovers could be stronger for BoP, where standardization and integrated system design can reduce customization across projects, than for EPC, where labor, permitting, and site-specific execution remain more deeply local. This creates a natural rationale for a hybrid learning structure in which BoP learning is treated as more global while EPC learning remains predominantly local.

## Quantitative implications of alternative learning structures

### Stacks

Having established several plausible learning structures for electrolyzer stack costs, we now examine their implications under a stylized deployment scenario. To isolate the effects of model structure, we consider a case in which global installed electrolysis capacity reaches between 60 GW and 140 GW and is evenly distributed across four technology variants: Western PEM, Chinese PEM, Western ALK, and Chinese ALK. This range is stylized but remains well below the announced global project pipeline of 520 GW as of 2024, of which only 20 GW had reached final investment decisions. Moreover, the equal allocation is not intended to represent the most likely market outcome, but to provide a transparent benchmark that isolates the effects of alternative learning structures from differences in deployment shares.

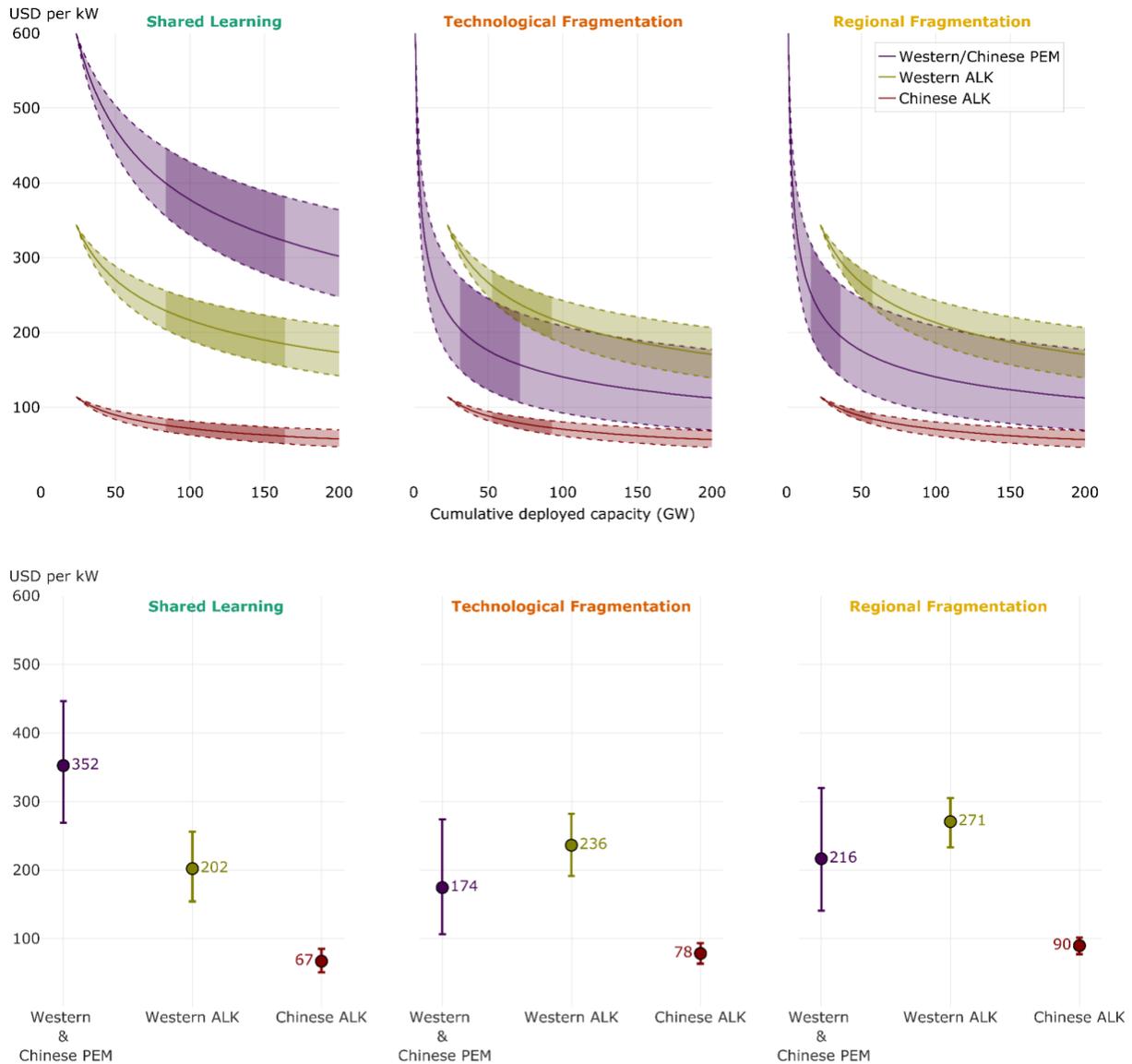

**Figure 1:** The top panel illustrates the evolution of electrolysis stack costs under three learning scenarios. The solid line represents the base case (20% learning rate), while the dashed lines reflect lower (15%) and higher (25%) learning rates. Initial costs are based on a recent Bloomberg New Energy Finance survey[36]. Shaded regions capture the range of projected costs based on a global installed electrolysis capacity between 60 and 140 GW, evenly distributed across the four technology variants. The bottom panel presents the resulting cost ranges across learning models.

We apply three learning structures: a shared learning model, in which all variants follow a common experience curve; a technology-fragmented model, in which ALK and PEM follow separate curves; and a regionally fragmented model, in which Western and Chinese variants also learn independently within each technology category. Unless otherwise noted, the base case assumes a learning rate of 20%, with sensitivities spanning 15% to 25%, consistent with observed learning rates for modular mass-manufactured technologies[39].

As shown in Figure 1, these alternative structures produce materially different stack cost projections. Under the shared learning model, both Western and Chinese PEM remain more expensive than Western ALK throughout the projection range. Under the fragmented learning models, by contrast, PEM eventually becomes less costly than Western ALK. The source of this divergence lies in the experience base from which each technology learns. In the shared model, PEM learning begins from a much larger aggregate installed base of roughly 24 GW, composed almost entirely of ALK deployment in the chlor-alkali industry. In the technology-fragmented model, PEM instead begins from its own installed base of approximately 1 GW. Because cost declines in experience-curve models occur through doublings of cumulative deployment, the smaller PEM baseline in the fragmented case allows more rapid early cost declines. When learning is further split between Western and Chinese variants, cumulative deployment is divided across still smaller learning families, reducing the scope for cost reduction within each one. The magnitude of these differences depends on both deployment scale and the assumed learning rate.

The divergence across learning structures is even clearer when expressed in terms of learning investment, defined here as the cumulative capital expenditure on stack procurement implied by the experience curve framework to reach a specified cost target. This metric should be interpreted as a model-implied quantity under the assumed deployment and learning structure, rather than as a literal forecast of future spending requirements. In the base case, reducing PEM stack costs from 600 USD/kW to 300 USD/kW would require approximately 60 billion USD of cumulative global investment in stack procurement under the shared learning model. Under the fragmented learning models, the same cost target is reached with only about 3 billion USD of cumulative stack investment and roughly 9 GW of deployment (see Figure 2). This difference arises because PEM moves through early doublings much faster when it learns from its own smaller installed base rather than from a much larger aggregate stack category. For more ambitious cost targets, the divergence across learning structures becomes even larger.

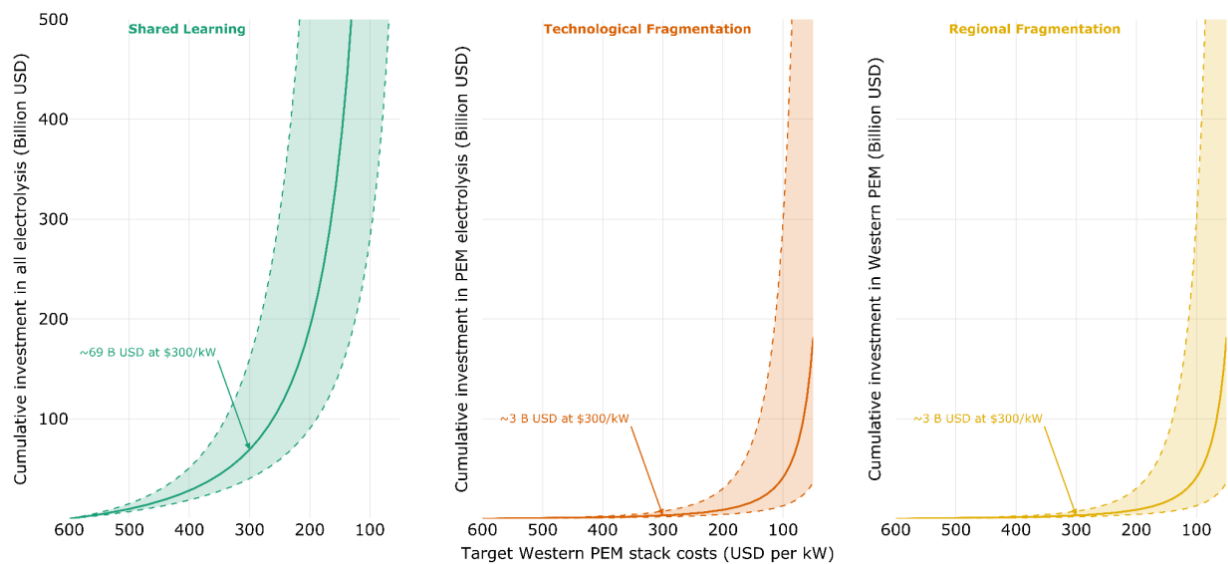

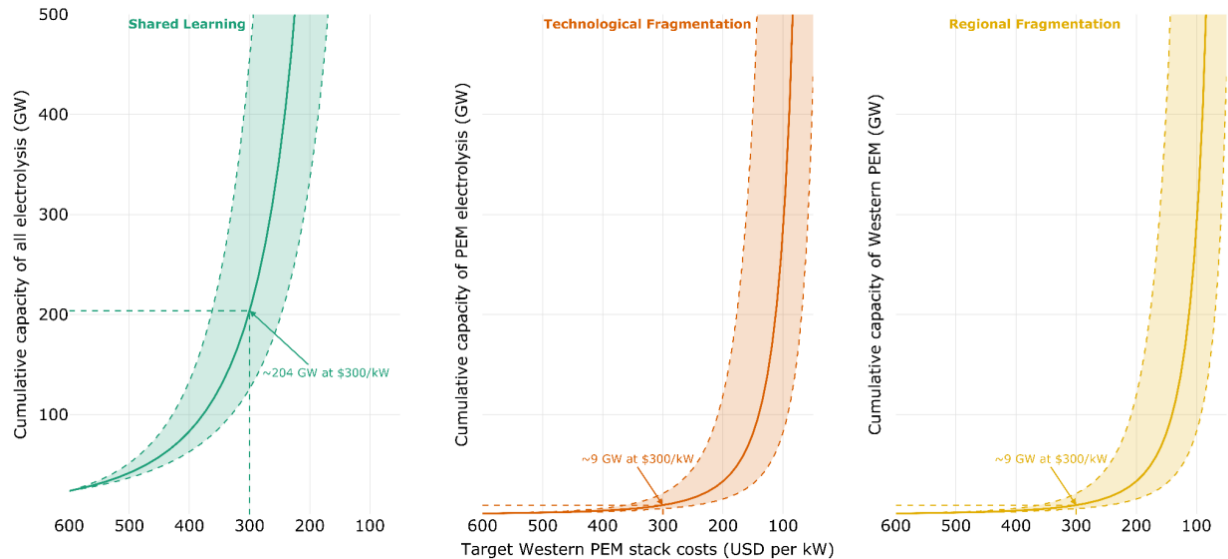

**Figure 2:** Learning investment (above) and learning capacity (below) required to reach various Western PEM electrolysis stack cost targets. The continuous line corresponds to the base (20%) learning rate, whereas the dashed lines correspond to low (15%) and high (25%) learning rates.

This pattern is especially visible in the case of reducing Western PEM stack costs to 100 USD/kW, a level comparable to the current cost of Chinese ALK stacks. Under the fragmented learning model, this target is reached with roughly 41 billion USD of cumulative investment and 287 GW of deployment. Under the shared learning, by contrast, the same target would require more than 890 billion USD of cumulative investment and 6180 GW of deployment. These differences imply very different views of the plausibility of deep cost reductions. In the fragmented learning models, a target such as 100 USD/kW remains demanding but is reached at scales that are at least within the range of the current project pipeline. In the shared model, the same target appears far less attainable because the aggregate learning base is so large that much more cumulative deployment is required to generate additional doublings.

These disparities reflect how each learning structure allocates experience. The shared learning model pools cumulative deployment across all stack variants. The technology-fragmented model isolates learning within PEM and ALK separately. The regionally fragmented model further divides learning between Western and Chinese variants within each technology. As a result, in a world with regionalized supply chains and limited cross-border knowledge diffusion, achieving a given cost target in all regions requires substantially more aggregate deployment and investment than in a globally integrated system. In that sense, the fragmented models provide a stylized quantification of the economic consequences of supply-chain fragmentation and barriers to the diffusion of manufacturing knowledge. More broadly, these results show that projected stack costs and implied learning investments are highly sensitive to how learning is structured, even when the assumed learning rate is unchanged.

Differences in stack-cost projections also translate into different estimates of hydrogen production cost. Figure 3 shows the contribution of stack costs to the levelized cost of hydrogen (LCOH) across utilization rates. Higher utilization reduces the effect of model structure on LCOH because differences in capital cost are spread over a larger volume of hydrogen output. Even so, the structural uncertainty remains meaningful. For a project operating at a 50% utilization rate (i.e. representing a project at powered by good wind resources), the difference across learning structures amounts to approximately 0.34 USD/kg for Western and Chinese PEM, 0.12 USD/kg for Western ALK, and 0.05 USD/kg for Chinese ALK. This difference is economically meaningful; for context, hydrogen produced from natural gas in the United States is often estimated to cost around 1 USD/kg due to low natural gas prices[40].

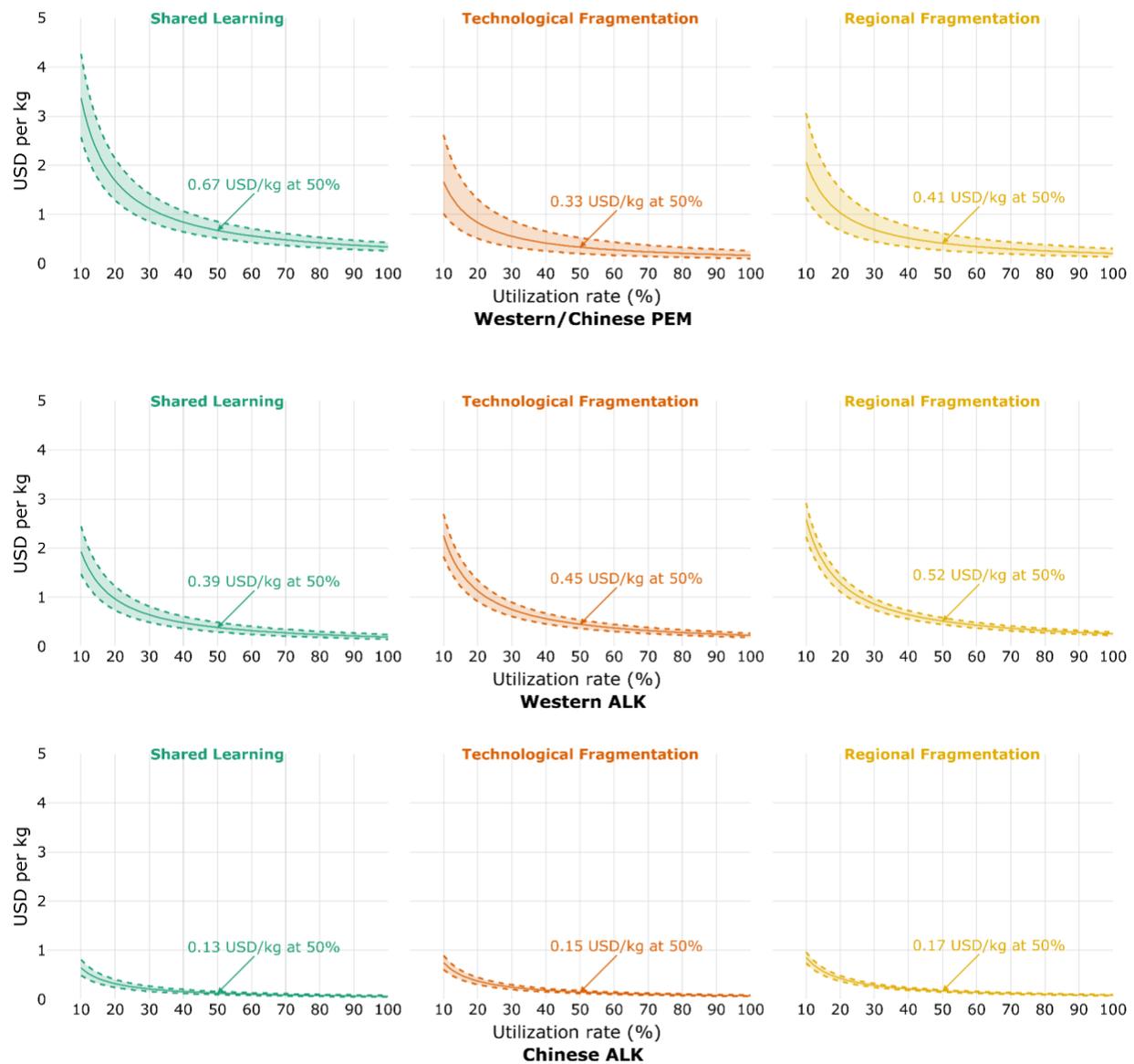

**Figure 3:** Projected contribution of stack costs to LCOH as a function of electrolysis utilization rate across different learning models. The dashed lines correspond to low (15%) and high (25%) learning rates. The assumed weighted average cost of capital is 7.6%[41]. Total cost of hydrogen production also includes contributions from balance of plant, engineering, procurement and construction costs, and costs of input electricity not included here.

## BoP and EPC

We next examine the implications of local versus global learning for balance of plant and engineering, procurement, and construction costs across four regions: the United States, Europe, China, and the rest of the world. We compare three stylized learning structures: fully local learning for both BoP and EPC, fully global learning for both, and a hybrid structure in which BoP learning is global while EPC learning remains local. The regional aggregation is necessarily coarse, especially for the rest of the world, but is constrained by the availability of regional cost data. To parameterize the exercise, we use a BloombergNEF forecast for regional electrolysis deployment in 2030, according to which the US, EU, China, and the rest of the world deploy 10 GW, 36 GW, 27 GW, and 25 GW, respectively. As a base case, we assume a 10% learning rate for both BoP and EPC in all regions, consistent with the expectation that these soft-cost components learn more slowly than core modular equipment. However, evidence from the solar industry suggests that BoP and EPC components may exhibit differing learning rates across different regions[42]. These assumptions are again intended to illustrate the implications of alternative learning structures rather than to provide a definitive forecast.

Figure 4 shows that the effects of model structure differ substantially across regions. The divergence between local and global learning is greatest where initial costs are high and where the gap between local and global deployment experience is large. The hybrid case lies between the fully local and fully global cases. In the United States, for example, the difference between a fully localized and a fully globalized learning structure amounts to approximately 215 USD/kW under the base-case deployment scenario. In China, by contrast, the difference is negligible because initial BoP and EPC costs are already relatively low. For the US, this divergence translates into a difference of about 0.41 USD/kg in hydrogen production cost for a project operating at a 50% utilization rate, as shown in Figure 5. Structural assumptions about BoP and EPC learning can therefore materially influence regional hydrogen cost estimates. Considered jointly with the differences arising from alternative stack learning structures, the implied difference in hydrogen cost can reach roughly 0.75 USD/kg in the US case. This is on the order of three quarters of the estimated cost of natural gas-based hydrogen in the United States.

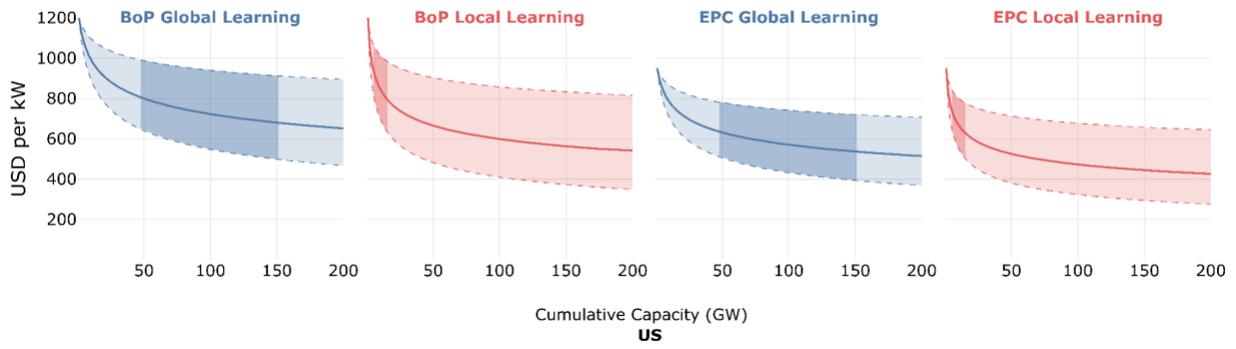
**US**

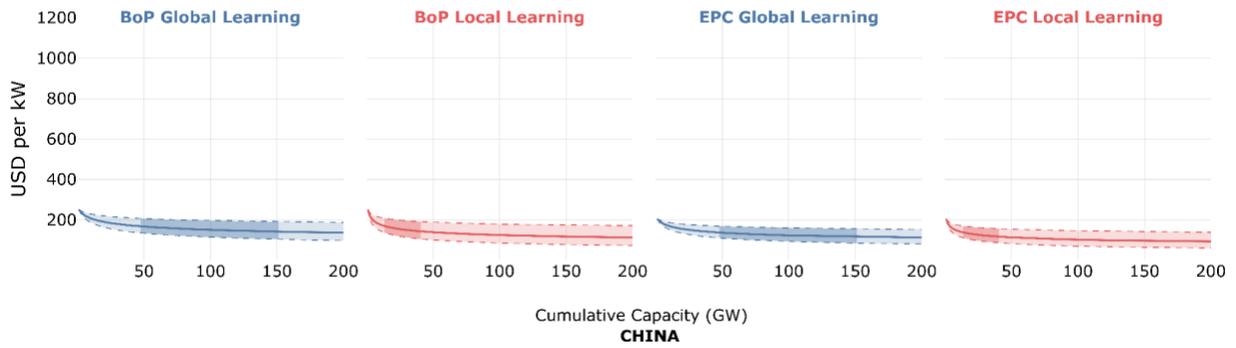
**CHINA**

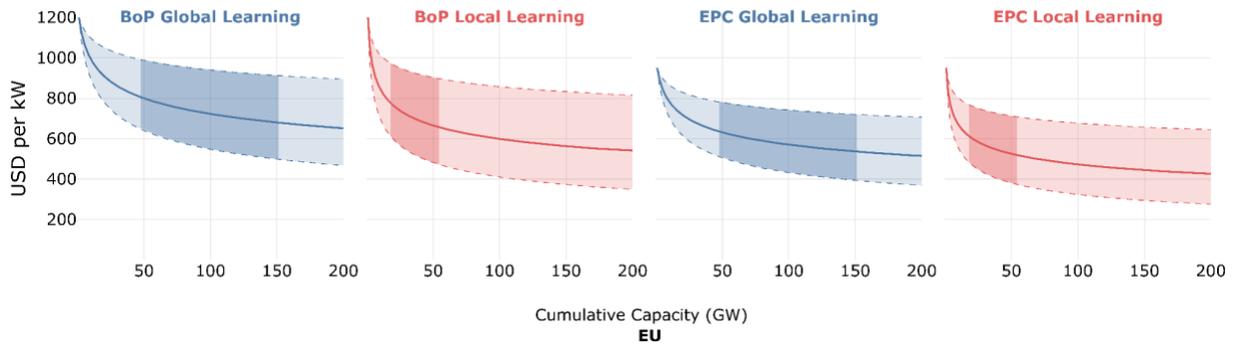
**EU**

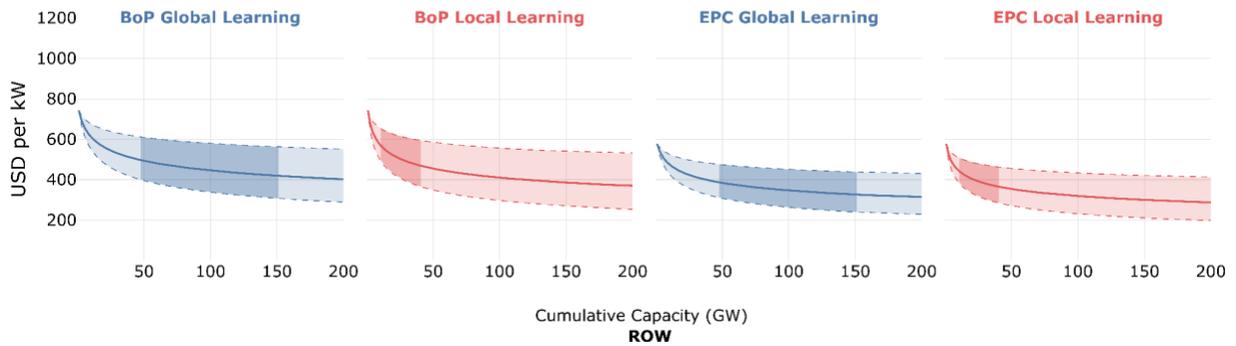
**ROW**

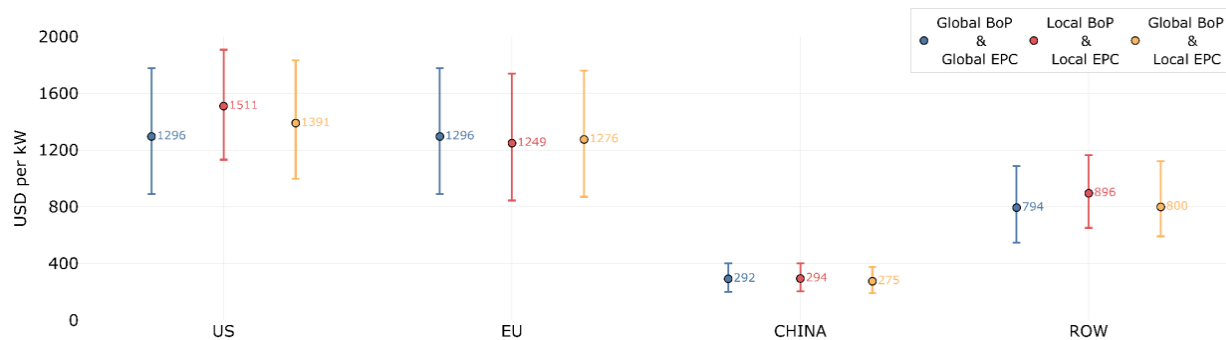

**Figure 4:** BoP and EPC learning curves across different learning models and regions, and projected 2030 costs using BNEF projections. The shaded area and uncertainty bars indicate uncertainty over deployed capacities (varied by 50% relative to the base case) and low (5%) and high (15%) learning rates. Initial costs are based on a recent Bloomberg New Energy Finance survey[36].

The implications are also large when expressed in terms of model-implied learning investment and capacity expansion. Under fully localized learning, the United States would need to deploy roughly 54 GW of electrolysis capacity and invest about 64 billion USD to reduce BoP and EPC costs to 1000 USD/kW, excluding stack costs. Under a fully global learning structure, achieving the same cost target would require approximately 189 GW of cumulative global deployment and 224 billion USD of global investment. However, it is important to note that the global investment requirement does not directly translate into a proportional investment burden for the US. Under the global learning model, the United States would still benefit from the cost reductions facilitated by international deployment, regardless of domestic deployment. This difference highlights why the local-versus-global learning assumption is consequential not only for projected costs, but also for questions of industrial strategy, first-mover advantage, and the extent to which deployment in one region can lower costs for others.

Taken together, these results show that structural uncertainty in learning models is not a secondary modeling detail, but a first-order determinant of projected electrolysis costs, implied investment requirements, and hydrogen cost estimates. At the same time, the results are derived from stylized deployment and learning assumptions chosen to isolate the implications of alternative learning structures. The accompanying dashboard extends this benchmark analysis by allowing users to explore alternative deployment pathways and parameter choices.

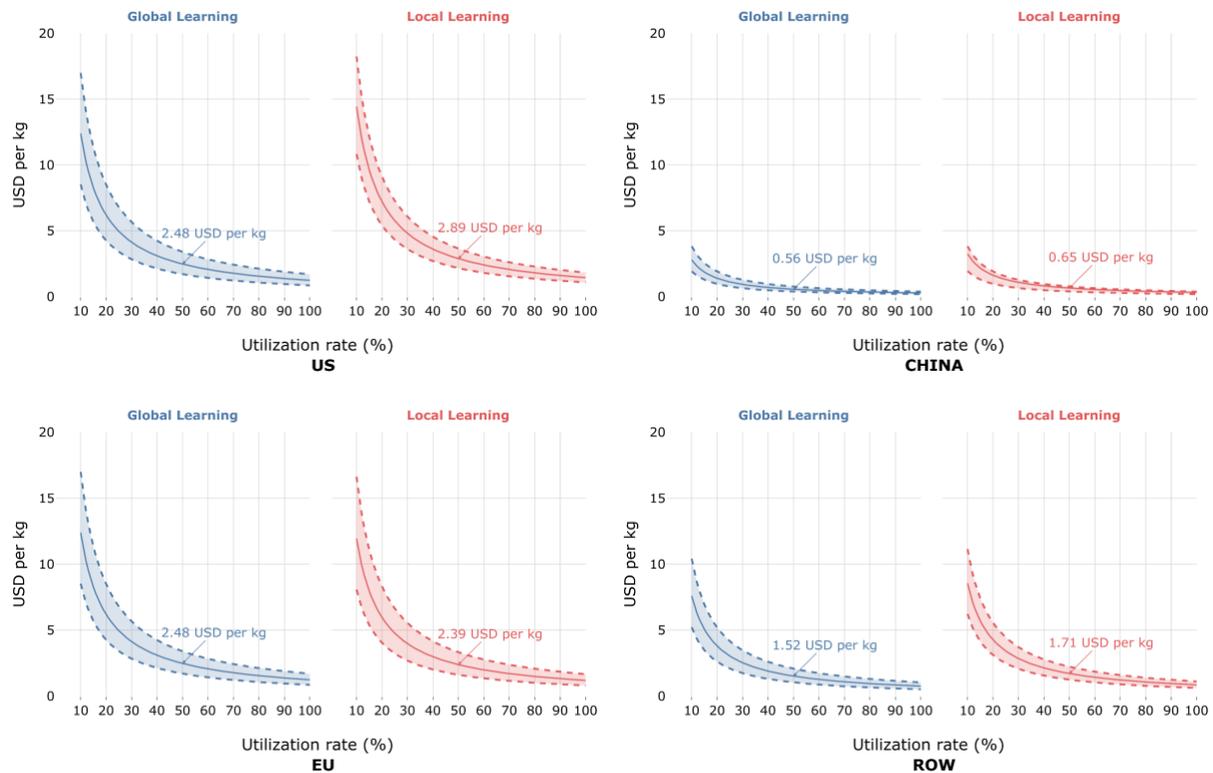

**Figure 5:** Projected contribution of BoP and EPC costs to LCOH in 2030 across different learning models. The shaded area indicates uncertainty over deployed capacities (varied by 50% relative to the base case) and low (5%) and high (15%) learning rates.

# Discussion

## Structural uncertainty is a first-order modeling issue

Forecasting the cost evolution of emerging technologies is deeply uncertain not only because learning rates are difficult to estimate in the early stages of deployment, but also because the structure of the learning model itself is uncertain. As shown here for electrolyzer stacks, balance-of-plant and EPC costs, alternative, defensible learning structures can generate materially different projections even when the assumed learning rate is unchanged. Widely used unified global learning models are therefore not neutral simplifications; they embed strong assumptions about spillovers across technology variants, regions, and cost components. Model structure should accordingly be treated as a first-order modeling choice, deserving as much scrutiny as the selection of learning-rate parameters.

No single learning model can be treated as universally correct. The appropriate structure depends on the technology, the nature of spillovers across variants and regions, the composition of project costs, and the policy or strategic question being asked. For that reason, experience-curve analysis should not rely on a single default specification when informing national strategy, energy-system modeling, or business planning. Interactive tools such as the online dashboard developed here for electrolysis are useful precisely because they allow users to explore the implications of alternative structural assumptions, parameter choices, and deployment pathways rather than treating any one specification as definitive.

## Implications for hydrogen strategy and industrial policy

These findings carry direct implications for national hydrogen strategy. Policymakers setting deployment targets should evaluate them under multiple learning structures rather than a single global learning curve, particularly when those targets are justified partly by the expectation that scale-up will unlock rapid future cost reductions[43]. If learning is more localized than commonly assumed, the scale of deployment required to achieve a given cost target may be much larger domestically, and the feasibility of existing strategy targets may need to be reassessed accordingly. This is especially important for countries seeking to position themselves as exporters of hydrogen or hydrogen-derived products such as ammonia and steel, since local learning can generate first-mover advantages and limit the ability of later entrants to benefit fully from deployment elsewhere.

The results also speak to industrial policy more broadly. Policies such as domestic manufacturing mandates, local-content requirements, export restrictions on manufacturing equipment, or limits on cross-border investment and collaboration do not merely affect where production occurs; they may also alter the structure of learning itself. Fragmented supply chains may support domestic capability formation and strategic autonomy, but they may also slow aggregate cost decline by reducing spillovers across regions. The relevant policy tradeoff is therefore not simply between domestic production and imports, but between alternative learning environments: one in which knowledge, design improvements, and manufacturing capability diffuse relatively broadly, and one in which they are increasingly trapped within parallel regional systems. More fundamentally, the structure of learning should not be treated as a fixed background condition. Policy choices can shape whether learning is accelerated, impeded, fragmented, or unified.

## Choosing among learning models

A central implication of this study is that the choice of learning model should be guided by the technological, industrial and policy reality the analysis is intended to capture. The appropriate learning structure depends on the technology's physical and manufacturing characteristics, the degree of spillover across variants and regions, and the decision context in which the model is being used. In practice, multiple structures may be defensible for the same technology. The table below provides guidance on the conditions under which different learning structures may be most appropriate and the kinds of policy or strategic questions they are best suited to address. The accompanying dashboard complements this guidance by enabling users to examine how alternative deployment pathways, parameter choices, and structural assumptions interact in shaping projected outcomes.

| Learning structure | When it is most appropriate | Core assumption | Best suited questions | Main risk if misapplied |
|---|---|---|---|---|
| **Shared global learning** | The technology is relatively standardized, supply chains are integrated, and knowledge diffusion across firms, variants, and regions is expected to be strong | Deployment of a variant anywhere generates broad spillovers that reduce costs elsewhere | How quickly might costs fall globally under open trade and broad diffusion of innovation? What global investments are needed to reach specific cost targets? | May understate domestic capability requirements and the consequences of fragmented supply chains |
| **Technology fragmented learning** | Competing variants differ substantially in materials, design, supply chains and manufacturing processes | Deployment of one variant provides limited spillovers to another, so each follows its own learning trajectory | Should deployment be concentrated on one variant or split across several? How should competition between different variants be evaluated? | May overstate separation if spillovers across variants are stronger than assumed |
| **Regionally fragmented learning** | Supply chains are increasingly localized, protectionist policies are important, or barriers to cross-border knowledge diffusion are significant | Learning is partly retained within regional manufacturing ecosystems rather than diffusing globally | What are the cost implications of techno-nationalist industrial policy, diffusion barriers (e.g. export controls on manufacturing equipment), or fragmented supply chains? | May overstate fragmentation if local manufacturing still allows cross-regional know-how diffusion. For instance, through foreign firms establishing local production |
| **Fully local learning for BoP and EPC** | Costs are dominated by labor, permitting, regulation, local market conditions, and domestic project-development experience | Cost reductions depend primarily on deployment experience gained within each region | How much domestic deployment is needed to reduce project costs locally? What first-mover advantages arise from local capability formation? | May understate the role of standardization, international developers diffusing project design knowledge and learning in multiple jurisdictions |
| **Hybrid learning: global BoP, local EPC** | Standardized equipment packages and integrated plant designs can diffuse internationally, but project execution remains location-specific | Some project costs benefit from global learning, while EPC continues to depend mainly on regional deployment experience | What cost reductions could be achieved by pushing standardization beyond core equipment to integrated plant designs? | May understate how plant-level standardization links BoP and EPC cost reductions across regions |

**Additional uncertainties and research directions**

This analysis also highlights several unresolved questions. One is whether fragmentation affects not only the structure of learning, but also the learning rate itself. If learning becomes increasingly separated across manufacturing ecosystems, then different regions may exhibit systematically different rates of cost decline. China's dense supplier networks, manufacturing capabilities, and organizational scale have contributed to rapid cost reductions in solar photovoltaics and lithium-ion batteries[17,44–46], and similar dynamics may influence electrolysis. In that case, competition between regions would depend not only on cumulative deployment, but also on differences in the pace of learning. Although the present study does not model regional heterogeneity in learning rates explicitly, this possibility reinforces the need for more research on how fragmented supply chains and industrial ecosystems could shape cost trajectories in emerging technologies.

Related uncertainties apply to balance-of-plant and EPC costs. This study assumes common BoP and EPC learning rates across regions, but evidence from the solar industry suggests that soft-cost learning may also differ geographically[42]. Future work should therefore examine how policy, market structure, developer experience, permitting, and local institutional conditions influence learning rates in these cost categories, and how those factors interact with model structure.

Technology design and industry organization may also reshape the structure of learning itself. Integrated offerings that combine stacks and balance-of-system components, as well as closer partnerships between technology providers and EPC firms, may increase standardization and expand the share of project costs that can benefit from broader global learning. An important open question is how strongly these design and organizational choices alter the boundary between globally and locally learned cost components, and whether policy can help move more of the cost structure into categories that benefit from broader learning spillovers.

## Implications for global decarbonization

The structure of technological learning also has important implications for climate policy and decarbonization in emerging economies. Many mitigation strategies implicitly rely on a global learning logic in which early deployment in wealthier countries lowers costs and enables later adoption elsewhere. If learning is more localized than this view assumes, that logic weakens. Cost reductions in emerging economies would then depend more heavily on earlier domestic deployment, supply-chain development, and institutional learning. In such a setting, countries would need to build local capabilities while technologies remain relatively expensive, compounding the challenge posed by high costs of capital and other structural barriers to clean-technology deployment[47]. The extent to which learning is global or local therefore shapes not only industrial advantage, but also the plausibility of broader global diffusion pathways for decarbonization.

More fundamentally, the structure of learning should not be viewed as a foregone conclusion. It is shaped in part by policy, industrial organization, and technology design. Policies that restrict trade, knowledge diffusion, investment, or labor mobility may fragment learning across regions and supply chains, while policies that encourage standardization, interoperability, integrated system design, and broader diffusion of manufacturing and project-development knowledge may increase the share of costs subject to wider learning spillovers. In this sense, policy can do more than support the deployment of emerging technologies; it can also influence whether learning is localized or broadly shared, and therefore how quickly cost reductions diffuse across regions. This suggests that cost-reduction strategies should adopt a more holistic view of innovation. Traditional emphasis on core components that are mass-produced and more likely to benefit from global learning, such as solar modules, battery cells, and electrolyzer stacks, may be insufficient where a large share of total cost lies in plant-level systems and soft costs. Standardization and integration of complete plant designs and technology packages may therefore offer one way to increase the share of costs exposed to broader learning and to support wider international diffusion.

**Broader applicability of the framework**

Although this study focuses on electrolysis, the framework is likely relevant to other emerging technologies that share analogous modeling challenges, including multiple technological variants, a mix of mass manufactured more standardized core components and project-specific costs, and uncertainty over whether learning should be treated as global or local. This is likely to be the case for other strategically important technologies such as small modular nuclear reactors. The significance of the issues raised here therefore extends beyond electrolysis itself: for a broader class of emerging technologies, projected costs may depend not only on learning rates, but also on how learning is structured across variants, components, and regions. This in turn raises additional strategic questions, including how strongly learning spills over across competing variants and how investment should be allocated across multiple variants under uncertainty. Concentrating investment in a single variant may maximize learning-by-doing and accelerate cost reduction, but diversification across several variants may reduce technology risk even as it fragments deployment and slows learning within each pathway. Effective technology policy therefore requires not only support for scale-up, but also careful attention to how scale-up is distributed across competing variants and how that distribution interacts with learning.

# Conclusion

This study shows that uncertainty in experience-curve analysis for emerging technologies does not arise only from poorly understood learning rates. It also arises from uncertainty over the structure of learning itself. In the case of electrolysis, alternative, defensible learning structures generate materially different projections of stack costs, balance-of-plant and EPC costs, hydrogen production costs, and the cumulative investment and deployment required to reach cost targets. Structural uncertainty is therefore not a secondary modeling detail, but a first-order determinant of the conclusions drawn from technological cost forecasts.

This issue is particularly important for technologies such as electrolysis, which have attracted substantial policy support and investment, often on the premise that such intervention will help drive down hydrogen production costs over time. As the electrolysis case shows, a higher degree of learning fragmentation reduces the effectiveness of each dollar invested, whether public or private, in driving cost reductions, even if it may also create first-mover advantages for some regions. No single learning model can be treated as universally correct, but the common assumption of a unified global learning curve should not be applied uncritically. Instead, it is essential to carefully assess the conditions and contexts that might justify one learning model rather than another and, ideally, to integrate findings from multiple learning models to enhance the robustness of cost projections, improve decision-making, and develop more informed strategies for technological deployment and investment.

# Methods

### Stack learning models

To project electrolyzer stack costs, we compare three alternative learning structures: shared learning, technology-fragmented learning, and regionally fragmented learning. In all cases, projected costs are calculated using a standard one-factor experience curve:

$$C_{s,\ell}^{\text{proj}} = C_{s,\ell}^{\text{cur}} \Big( \sum_{v \in \mathcal{F}_{s,\ell}} x_v^{\text{proj}} \Big/ \sum_{v \in \mathcal{F}_{s,\ell}} x_v^{\text{cur}} \Big)^{\log(1 - LR_s)/\log 2}$$

where $C_{s,\ell}^{\text{cur}}$ and $C_{s,\ell}^{\text{proj}}$ denote the current and projected costs of stack variant $s$ under learning structure $\ell$; $x_v^{\text{cur}}$ and $x_v^{\text{proj}}$ denote the current and projected cumulative deployed capacities of variant $v$; $\mathcal{F}_{s,\ell}$ denotes the learning family associated with stack variant $s$ under learning structure $\ell$; and $LR_s$ is the stack learning rate.

$$s \in \{\text{Western ALK, Chinese ALK, Western PEM, Chinese PEM}\}$$

$$\ell \in \{\text{shared, technology} - \text{fragmented, regionally} - \text{fragmented}\}$$

The learning families differ across the three cases. Under shared learning, all four stack variants contribute to a common experience base:

$$\mathcal{F}_{s,\text{shared}} = \{\text{Western ALK, Chinese ALK, Western PEM, Chinese PEM}\}$$

Under technology-fragmented learning, ALK variants learn together and PEM variants learn together:

$$\mathcal{F}_{s,\text{technology-fragmented}} = \{\text{Western ALK, Chinese ALK}\}, \quad s \in \{\text{Western ALK, Chinese ALK}\}$$

$$\mathcal{F}_{s,\text{technology-fragmented}} = \{\text{Western PEM, Chinese PEM}\}, \quad s \in \{\text{Western PEM, Chinese PEM}\}$$

Under regionally fragmented learning, each regional stack variant follows its own experience curve independently:

$$\mathcal{F}_{s,\text{regionally-fragmented}} = \{s\}$$

The differences across these models therefore arise entirely from how cumulative deployment is allocated across experience bases as well as the assumed learning starting point.

## BoP and EPC learning models

To project balance-of-plant (BoP) and engineering, procurement, and construction (EPC) costs, we compare three alternative learning structures: fully local learning, fully global learning, and a hybrid structure in which BoP learning is global while EPC learning remains local. For a given cost category $k$ in region $r$, projected costs are calculated as:

$$C_{k,r,\ell}^{\text{proj}} = C_{k,r,\ell}^{\text{cur}} \Big( \sum_{q \in \mathcal{G}_{r,\ell}} x_q^{\text{proj}} \Big/ \sum_{q \in \mathcal{G}_{r,\ell}} x_q^{\text{cur}} \Big)^{\log(1 - LR_{k,r})/\log 2}$$

where $C_{k,r,\ell}^{\text{cur}}$ and $C_{k,r,\ell}^{\text{proj}}$ denote the current and projected costs of category $k$ in region $r$ under learning structure $\ell$; $x_q^{\text{cur}}$ and $x_q^{\text{proj}}$ denote the current and projected cumulative deployed electrolysis capacities in region $q$; $\mathcal{G}_{r,\ell}$ denotes the learning family associated with region $r$ under learning structure $\ell$; and $LR_{k,r}$ is the learning rate for cost category $k$ in region $r$. In the benchmark analysis in the main text, BoP and EPC learning rates are assumed to be uniform across regions, although the notation allows for regional variation.

$$k \in \{\text{BoP, EPC}\}, \quad r \in \{\text{US, EU, China, ROW}\}$$

$$\ell \in \{\text{local, global, hybrid}\}$$

Under fully global learning, all regions contribute to a common experience base:

$$\mathcal{G}_{r,\text{global}} = \{\text{US, EU, China, ROW}\}$$

Under fully local learning, each region learns only from its own deployment:

$$\mathcal{G}_{r,\text{local}} = \{r\}$$

In the hybrid case, BoP costs are modeled using global learning, while EPC costs are modeled using local learning:

$$\mathcal{G}_{r,\text{hybrid}}^{\text{BoP}} = \{\text{US, EU, China, ROW}\}$$

$$\mathcal{G}_{r,\text{hybrid}}^{\text{EPC}} = \{r\}$$

This formulation allows the model to distinguish between cost categories that may benefit from broader international spillovers, such as standardized and integrated system designs, and those that may remain more strongly tied to local deployment experience, such as EPC and project execution.

## Data Availability

Cost and deployment projection data used in this study were obtained from Bloomberg New Energy Finance (BloombergNEF). All other inputs, including learning rate assumptions and model parameters, are reported in the main manuscript. The interactive dashboard accompanying this study allows full exploration of all model parameters and assumptions.

## Acknowledgments

This work was funded by Princeton University's Low-Carbon Technology Consortium, which is supported by unrestricted gifts from Google, GE, and ClearPath.

## Author contributions

Conceptualization, M.A. and J.D.J.; methodology,  M.A. and J.D.J; Investigation, M.A.; writing – original draft, M.A. ; writing – review & editing, J.D.J.; funding acquisition, J.D.J.; supervision, J.D.J.

## Declaration of interests

Jesse D. Jenkins is chief scientist of Resilient Transition, which provides modeling, analytics and decision support for utilities, investors, asset owners, and technology companies. He serves on the advisory boards of Eavor Technologies Inc., a closed-loop geothermal technology company, Rondo Energy, a provider of high-temperature thermal energy storage and industrial decarbonization solutions, Dig Energy, a developer of low-cost drilling solutions for geothermal heating and cooling, and Karman Industries, a developer of high-efficiency industrial heat pumps, and he has an equity interest in each company. He also serves as a technical advisor to MUUS Climate Partners and Energy Impact Partners, both investors in early-stage climate technology companies.

## Declaration of generative AI and AI-assisted technologies in the writing process

During the preparation of this work, the author used ChatGPT and Claude in order to improve readability, language and figures. After using these tools/services, the author reviewed and edited the content as needed and takes full responsibility for the content of the publication.

## References

1. Wright, T. P. Factors Affecting the Cost of Airplanes. *J. Aeronaut. Sci.* **3**, 122–128 (1936).

2. Colpier, U. C. & Cornland, D. The economics of the combined cycle gas turbine—an experience curve analysis. *Energy Policy* **30**, 309–316 (2002).

3. de La Tour, A., Glachant, M. & Ménière, Y. Predicting the costs of photovoltaic solar modules in 2020 using experience curve models. *Energy* **62**, 341–348 (2013).

4. Junginger, M., Faaij, A. & Turkenburg, W. C. Global experience curves for wind farms. *Energy Policy* **33**, 133–150 (2005).

5. Rubin, E. S., Yeh, S., Antes, M., Berkenpas, M. & Davison, J. Use of experience curves to estimate the future cost of power plants with CO2 capture. *Int. J. Greenh. Gas Control* **1**, 188–197 (2007).

6. Gan, P. Y. & Li, Z. Quantitative study on long term global solar photovoltaic market. *Renew. Sustain. Energy Rev.* **46**, 88–99 (2015).

7. Lindman, Å. & Söderholm, P. Wind power learning rates: A conceptual review and meta-analysis. *Energy Econ.* **34**, 754–761 (2012).

8. Mauleón, I. Photovoltaic learning rate estimation: Issues and implications. *Renew. Sustain. Energy Rev.* **65**, 507–524 (2016).

9. Neij, L. Use of experience curves to analyse the prospects for diffusion and adoption of renewable energy technology. *Energy Policy* **25**, 1099–1107 (1997).

10. Samadi, S. The experience curve theory and its application in the field of electricity generation technologies – A literature review. *Renew. Sustain. Energy Rev.* **82**, 2346–2364 (2018).

11. Schmidt, O., Hawkes, A., Gambhir, A. & Staffell, I. The future cost of electrical energy storage based on experience rates. *Nat. Energy* **2**, 1–8 (2017).

12. International Energy Agency. *Experience Curves for Energy Technology Policy*. (OECD, 2000). doi:10.1787/9789264182165-en.

13. Mayer, J. *et al. Current and Future Cost of Photovoltaics*. (2015) doi:10.13140/RG.2.1.1371.7206.

14. Haas, R., Sayer, M., Ajanovic, A. & Auer, H. Technological learning: Lessons learned on energy technologies. *WIREs Energy Environ.* **12**, e463 (2023).

15. Lovering, J. R., Yip, A. & Nordhaus, T. Historical construction costs of global nuclear power reactors. *Energy Policy* **91**, 371–382 (2016).

16. Huenteler, J., Schmidt, T. S., Ossenbrink, J. & Hoffmann, V. H. Technology life-cycles in the energy sector — Technological characteristics and the role of deployment for innovation. *Technol. Forecast. Soc. Change* **104**, 102–121 (2016).

17. Nemet, G. *How Solar Energy Became Cheap: A Model for Low-Carbon Innovation*.

18. Kavlak, G., McNerney, J. & Trancik, J. E. Evaluating the causes of cost reduction in photovoltaic modules. *Energy Policy* **123**, 700–710 (2018).

19. Klemun, M. M., Kavlak, G., McNerney, J. & Trancik, J. E. Mechanisms of hardware and soft technology evolution and the implications for solar energy cost trends. *Nat. Energy* **8**, 827–838 (2023).

20. Huenteler, J., Niebuhr, C. & Schmidt, T. S. The effect of local and global learning on the cost of renewable energy in developing countries. *J. Clean. Prod.* **128**, 6–21 (2016).

21. Yeh, S. & Rubin, E. S. A review of uncertainties in technology experience curves. *Energy Econ.* **34**, 762–771 (2012).

22. IRENA (2020), Green Hydrogen Cost Reduction: Scaling up Electrolysers to Meet the 1.5$^0$C Climate Goal, International Renewable Energy Agency, Abu Dhabi.

23. Schoots, K., Ferioli, F., Kramer, G. & Vanderzwaan, B. Learning curves for hydrogen production technology: An assessment of observed cost reductions. *Int. J. Hydrog. Energy* **33**, 2630–2645 (2008).

24. Bühler, L. & Möst, D. Derivation of one- and two-factor experience curves for electrolysis technologies. *Int. J. Hydrog. Energy* **89**, 105–116 (2024).

25. Glenk, G., Meier, R. & Reichelstein, S. Cost Dynamics of Clean Energy Technologies. *Schmalenbach J. Bus. Res.* **73**, 179–206 (2021).

26. Revinova, S., Lazanyuk, I., Ratner, S. & Gomonov, K. Forecasting Development of Green Hydrogen Production Technologies Using Component-Based Learning Curves. *Energies* **16**, (2023).

27. Reksten, A. H., Thomassen, M. S., Møller-Holst, S. & Sundseth, K. Projecting the future cost of PEM and alkaline water electrolysers; a CAPEX model including electrolyser plant size and technology development. *Int. J. Hydrog. Energy* **47**, 38106–38113 (2022).

28. Glenk, G., Holler, P. & Reichelstein, S. Advances in power-to-gas technologies: cost and conversion efficiency. *Energy Environ. Sci.* **16**, 6058–6070 (2023).

29. Böhm, H., Goers, S. & Zauner, A. Estimating future costs of power-to-gas – a component-based approach for technological learning. *Int. J. Hydrog. Energy* **44**, 30789–30805 (2019).

30. Atouife, M. & Jenkins, J. D. Learning-Driven Economics of Electrolytic Hydrogen Dashboard. https://learning-driven-economics-of-electrolytic-hydrogen-dashboard.streamlit.app/.

31. Ferioli, F., Schoots, K. & van der Zwaan, B. C. C. Use and limitations of learning curves for energy technology policy: A component-learning hypothesis. *Energy Policy* **37**, 2525–2535 (2009).

32. Krishnan, S., Corona, B., Kramer, G. J., Junginger, M. & Koning, V. Prospective LCA of alkaline and PEM electrolyser systems. *Int. J. Hydrog. Energy* **55**, 26–41 (2024).

33. The race to ramp up renewable green hydrogen goes high tech. https://www.siemens-energy.com/us/en/home/stories/electrolyzer-gigawatt-factory.html.

34. Watanabe, C., Wakabayashi, K. & Miyazawa, T. Industrial dynamism and the creation of a "virtuous cycle" between R&D, market growth and price reduction: The case of photovoltaic power generation (PV) development in Japan. *Technovation* **20**, 299–312 (2000).

35. Helveston, J. P., He, G. & Davidson, M. R. Quantifying the cost savings of global solar photovoltaic supply chains. *Nature* **612**, 83–87 (2022).

36. BloombergNEF. *Electrolyzer Price Survey 2024 Rising Costs, Glitchy Tech*. (2024).

37. High-power electrolyzers for the lowest cost clean hydrogen. *Electric Hydrogen* https://eh2.com/.

38. Worley. *FROM AMBITION TO REALITY 3*. https://www.worley.com/en/insights/our-thinking/energy-transition/from-ambition-to-reality/ (2022).

39. Malhotra, A. & Schmidt, T. S. Accelerating Low-Carbon Innovation. *Joule* **4**, 2259–2267 (2020).

40. White, J. B. (CONTR). Comparison of Commercial, State-of-the-Art, Fossil-Based Hydrogen Production Technologies.

41. Penev, M. *et al. Capital Structure for Techno-Economic Analysis of Hydrogen Projects*. NREL/TP--5400-90103, 2397248, MainId:91881 https://www.osti.gov/servlets/purl/2397248/ (2024) doi:10.2172/2397248.

42. Elshurafa, A. M., Albardi, S. R., Bigerna, S. & Bollino, C. A. Estimating the learning curve of solar PV balance–of–system for over 20 countries: Implications and policy recommendations. *J. Clean. Prod.* **196**, 122–134 (2018).

43. European Comission. *A Hydrogen Strategy for a Climate-Neutral Europe*. https://energy.ec.europa.eu/system/files/2020-07/hydrogen_strategy_0.pdf/ (2020).

44. Nahm, J. & Steinfeld, E. S. Scale-up Nation: China's Specialization in Innovative Manufacturing. *World Dev.* **54**, 288–300 (2014).

45. Helveston, J. & Nahm, J. China's key role in scaling low-carbon energy technologies. *Science* **366**, 794–796 (2019).

46. Nahm, J. China's Specialization in Innovative Manufacturing. in *Collaborative Advantage: Forging Green Industries in the New Global Economy* (ed. Nahm, J.) 0 (Oxford University Press, 2021). doi:10.1093/oso/9780197555361.003.0005.

47. IEA. *Cost of Capital Observatory*. https://www.iea.org/reports/cost-of-capital-observatory (2023).